\documentclass[letter]{aa} 
\newcommand{\HII}{H\,{\scriptsize II}}

\newcommand{\CII}{[C\,{\scriptsize II}]}

\newcommand{\OI}{[O\,{\scriptsize I}]}

\newcommand{\CI}{[C\,{\scriptsize I}]}
\newcommand{\water}{H$_2$O}

\newcommand{\vlsr}{v$_{\rm LSR}$}
\newcommand{\eleven}{11$\to$10}
\newcommand{\two}{2$\to$1}
\newcommand{\micron}{$\mu$m}
\newcommand{\kms}{km s$^{-1}$}
\newcommand{\kkms}{K km s$^{-1}$}
\newcommand{\scm}{cm$^{-2}$}
\newcommand{\ccm}{cm$^{-3}$}
\newcommand{\ergbrightness}{erg s$^{-1}$ cm$^{-2}$ sr$^{-1}$}

\newcommand{\thco}{$^{\rm 13}$CO}

\usepackage{graphicx}
\usepackage{natbib}
\usepackage{txfonts}
\begin{document}
%
\title{SOFIA observations of S106: Dynamics of the warm gas}


\author{
R. Simon\inst{1}
\and
N. Schneider\inst{2}
\and
J. Stutzki\inst{1}
\and
R. G\"usten\inst{3}
\and
U.~U. Graf\inst{1}
\and
P. Hartogh\inst{4}
\and
X. Guan\inst{1}
\and
J.~G. Staguhn\inst{5,6}
\and
D.~J. Benford\inst{5}
}
\institute{KOSMA, I. Physikalisches Institut, Universit\"at zu K\"oln, Z\"ulpicher Str. 77, 50937 K\"oln, Germany\\
\email{simonr@ph1.uni-koeln.de}
\and
Laboratoire AIM Paris Saclay, CEA/Irfu - Universit\'e Paris Diderot - CNRS, Centre d'\'etudes de Saclay, 91 191 Gif-sur-Yvette\\
\email{nicola.schneider-bontemps@cea.fr}
\and 
Max-Planck Institut f\"ur Radioastronomie, Auf dem H\"ugel 69, 53121 Bonn, Germany
\and
Max-Planck-Institut f\"ur Sonnensystemforschung, Max-Planck-Str. 2, 37191 Katlenburg-Lindau, Germany
\and
Observational Cosmology Laboratory (Code 665), NASA Goddard Space Flight Center, Greenbelt, MD 20771, USA
\and 
Department of Physics \& Astronomy, Johns Hopkins University, Baltimore, MD 21218, USA 
}
   \date{Received; accepted}
 \abstract
{The \HII\ region/PDR/molecular cloud complex S106 is excited by a
  single O-star. The full extent of the warm and dense gas close to
  the star has not been mapped in spectrally resolved high-J CO or
  \CII\ lines, so the kinematics of the warm, partially ionized gas,
  are unknown. 
  Whether the prominent dark lane bisecting the hourglass-shaped
  nebula is due solely to the shadow cast by a small disk around the
  exciting star or also to extinction in high column foreground gas
  was an open question until now.}
{To disentangle the morphology and kinematics of warm neutral and
  ionized gas close to the star, study their relation to the bulk of
  the molecular gas, and to investigate the nature of the dark lane.}
{We use the heterodyne receiver GREAT on board SOFIA to observe
  velocity resolved spectral lines of \CII\ and CO \eleven\ in
  comparison with so far unpublished submm continuum data at 350
  \micron\ (SHARC-II) and complementary molecular line
  data.}
{The high angular and spectral resolution observations show a very
  complex morphology and kinematics of the inner S106 region, with
  many different components at different excitation conditions
  contributing to the observed emission. The \CII\ lines are found to
  be bright and very broad, tracing high velocity gas close to the
  interface of molecular cloud and \HII\ region. CO \eleven\ emission
  is more confined, both spatially and in velocity, to the immediate
  surroundings of S106 IR showing the presence of warm, high density
  (clumpy) gas.  Our high angular resolution submm continuum
  observations rule out the scenario where the dark lane separating
  the two lobes is due solely to the shadow cast by a small disk
  close to the star. The lane is clearly seen also as warm, high
  column density gas at the boundary of the molecular cloud and
  \HII\ region.}
{}
\maketitle
%
\section{Introduction}
The \HII\ region S106 at a distance of 1.2 to 1.8 kpc
\citep{Schneider3} in Cygnus is a prominent bipolar emission nebula
associated with an extended molecular cloud.  The nebula is excited by
the single, late O-type star S106 IR, which creates a bright Photon
Dominated Region (PDR) at the molecular cloud interfaces. In addition,
S106 IR drives an ionized wind with a velocity of $\sim$200
\kms\ \citep{Simon1982} that is responsible for the hourglass shape
and high velocity wings seen in optically thick molecular line
emission close to the star. Here, we focus on the immediate
surroundings of S106 IR, relevant for the context of the SOFIA
observations. More details and references are given in
\citet{Hodapp2008} and by \citet{Schneider1,Schneider2,Schneider3}.

The two lobes of ionized gas seen in the optical and radio continuum
are separated by a dark lane that is very prominent in near- to mid-IR
imaging \citep{Smith2001,Oasa2006}.  (Sub)Millimeter dust and
optically thin molecular line emission
\citep{Richer1993,Schneider1,Schneider2} show two emission peaks
$\sim$$15''$ east and west of S106 IR in the dark lane, while the bulk
of the molecular gas is located further to the east and west of the
lobes. Two clusters of \water\ masers were detected at the western
peak \citep[S106 FIR,][]{Stutzki1982,Furuya1999}. \citet{Richer1993}
interpret this source as a Class 0 young stellar object.

While the dark lane was initially interpreted as a smooth, large-scale
disk \citep{Bally1982,Little1995}, \citet{Barsony1989} and
\citet{Richer1993} concluded that the molecular line and dust
continuum emission arises from a clumpy molecular cloud, possibly the
dense remnants of a large disk or torus disrupted by the central
star. A high degree of clumpiness was already inferred early on
through observations of ammonia \citep{Stutzki1985}.  Whether the dark
lane is just a shadow cast by a small, edge-on (accretion) disk close
to the star \citep{Bally1983,Noel2005}, largely confining the ionizing
radiation to the optical/radio lobes, or in addition due to dust
extinction in high column density gas, is a matter of ongoing
debate. A possible detection of the disk was reported by
\citet{Hoare1996} and \citet{Gibb2007}.


Observations in mid- to high-J CO lines, specifically tracing PDR gas
close to S106 IR, indicate the presence of warm (T$>$200 K) and dense
(n$>$10$^5$ \ccm) gas
\citep{Harris1987,Graf1993,Richer1993,Little1995,Schneider2}.  
\citet{Schneider2} observed a larger area (few arcmin) in \CII\ 158
$\mu$m using the Kuiper Airborne Observatory (KAO), and submm \CI\ and
CO lines (KOSMA 3 m), revealing the whole spatial extent of the PDR
region. Higher angular resolution \OI\ observations towards the center
provide a more detailed view of the higher density PDR gas. The
\CII\ and \OI\ KAO data, however, were not velocity resolved.

\section{Observations} \label{obs} 
The \CII\ atomic fine structure line at 1.900537 THz (157.737 $\mu$m)
and the CO \eleven\ rotational line at 1.267 THz were observed with
the heterodyne receiver GREAT\footnote{German Receiver for Astronomy
  at Terahertz. GREAT is a development by the MPI f\"ur
  Radioastronomie and the KOSMA/Universit\"at zu K\"oln, in
  cooperation with the MPI f\"ur Sonnensystemforschung and the DLR
  Institut f\"ur Planetenforschung.}  on board SOFIA during two flights
on April 13 and July 22, 2011, from Palmdale/California.
Total power on-the-fly maps with a scanning speed of 8$''$/s, 1~s dump
time, and a typical size of 2$'$$\times$2$'$ were
performed. Positional offsets refer to the star S106 IR
(R.A.,Dec.)(J2000)$=(20^h27^m26^s.74,37^\circ22'47''.9)$. The
region centered on S106 IR has three coverages while the western
extension has only one to two coverages, resulting in a non-uniform
noise distribution. Blank sky subtraction was done towards a position
offset by ($8'$,$0'$), well outside any emission seen in the KAO
\CII\ map presented by \citet{Schneider2}.
Instrument alignment and telescope efficiencies, antenna temperature
and atmospheric calibration, as well as the spectrometers used are
described in \citet{Heyminck2012} and \citet{Guan2012}. Here, we only
show data from one of the Fast Fourier Transform Spectrometers
(AFFTS), the other spectrometers giving redundant information. All
line intensities are reported as main beam temperatures scaled with
main-beam efficiencies of 0.51 and 0.54 for \CII\ and CO, and a
forward efficiency of 95\%. The mean r.m.s. noise temperatures per 0.5
\kms\ velocity bin for the central region (western extension) are 2.2
(3.7) K for \CII\ (16$''$ beam) and 1.8 (2.4) K for CO (19.6$''$
beam). The absolute calibration uncertainty is estimated to be
$\sim$10\%.  The IRAM 30 m observations used for comparison with the
new SOFIA data are described in \citet{Schneider1}.

The 350 \micron\ continuum observations were obtained in March, 2004,
with the SHARC-II bolometer camera at the CSO\footnote{The Caltech
  Submillimeter Observatory is operated by the California Institute of
  Technology under cooperative agreement with the National Science
  Foundation (AST-0838261).}. Our scanning strategy was to modulate
the telescope pointing with a non-connecting Lissajous pattern within
the limits of reasonable telescope acceleration (typical periods
of 20~s). Calibration was frequently verified on planets and
moons. Pointing was checked about every hour on evolved stars
and blazars. Intensities are given in Jy/beam with a beam size of
9$''$.

\begin{figure*}[th]
\includegraphics[height=5.75cm]{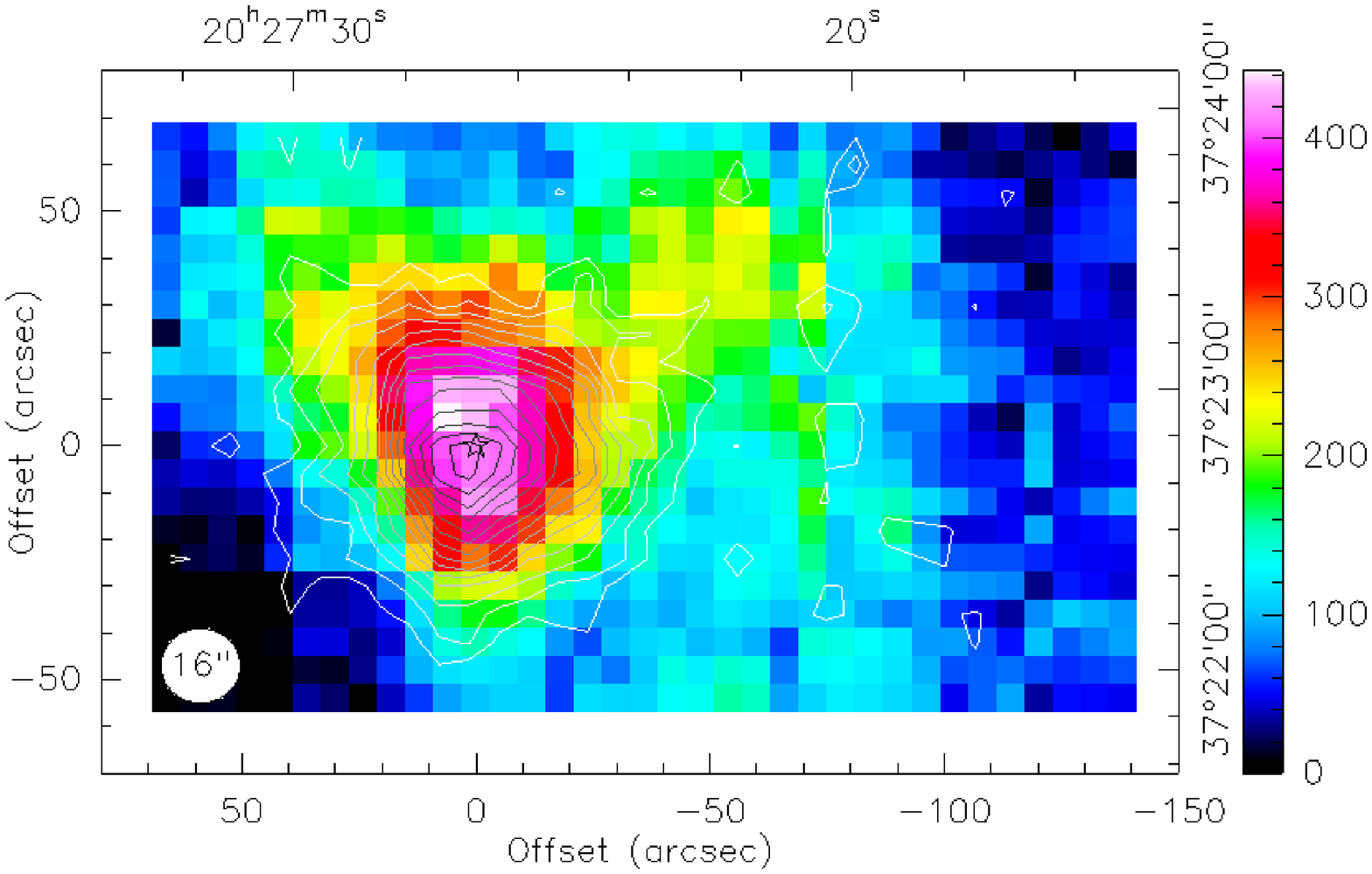}
\includegraphics[height=5.75cm]{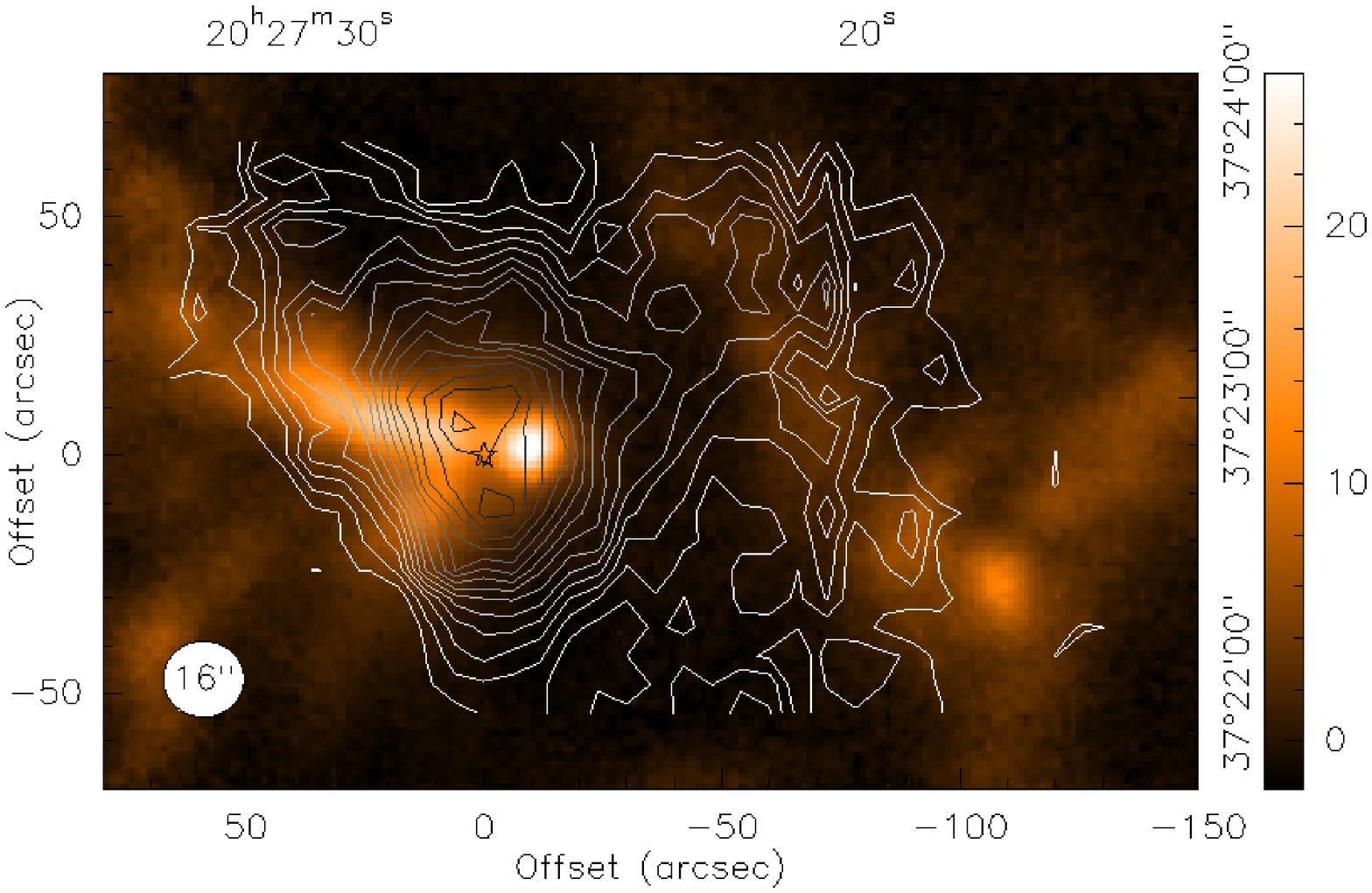}\\
\includegraphics[height=5.75cm]{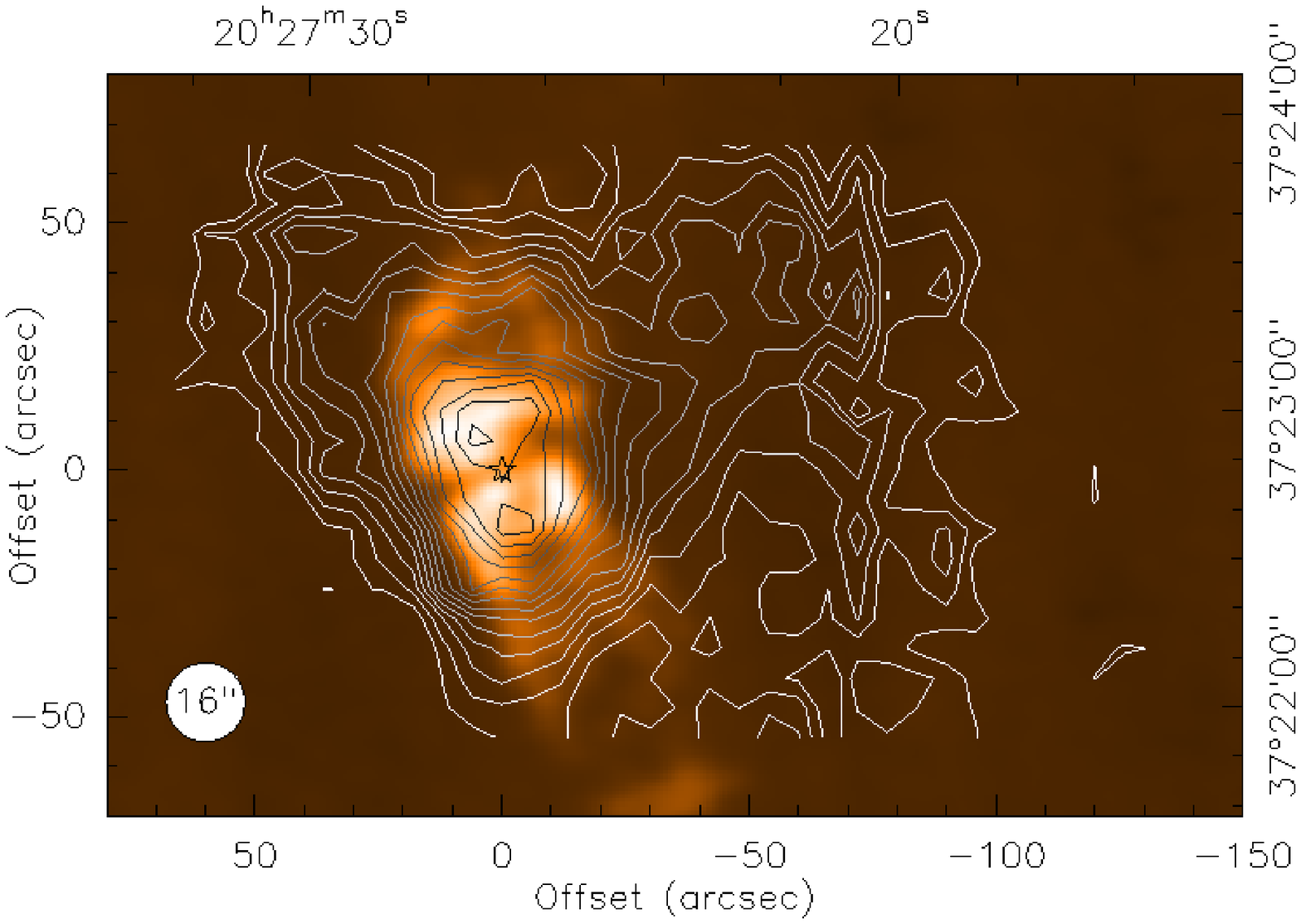} 
\hspace{8.5mm}
\includegraphics[height=5.75cm]{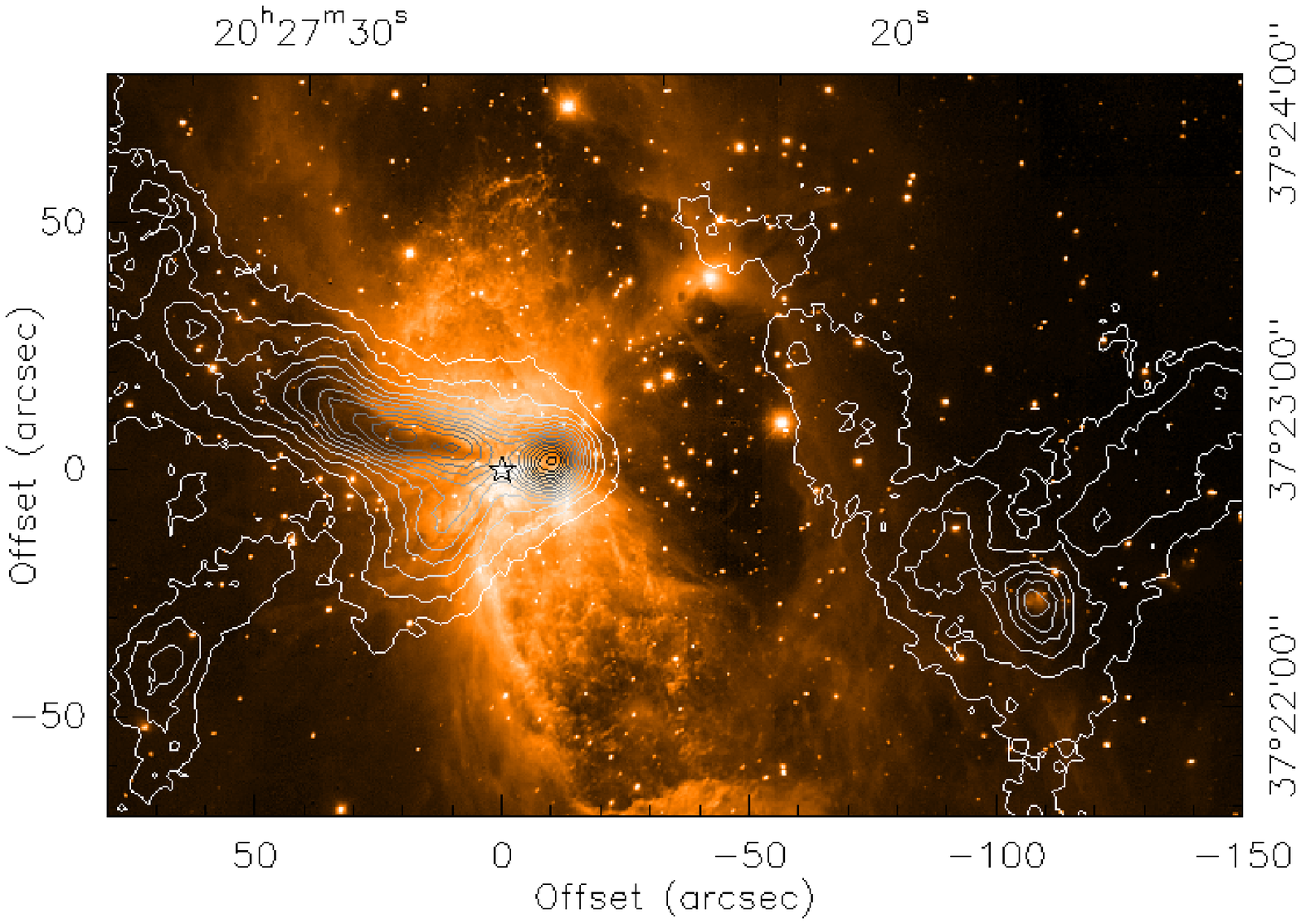}
\caption{Top: Integrated intensity (\vlsr\ --20 to 20 \kms) of
  \CII\ (color) and CO \eleven\ (contours from 15 to 215 in steps of
  15 \kkms). The position of S106 IR is indicated by a star and the
  SOFIA beam size for \CII\ is drawn in the bottom left corner. On the
  right, contours of \CII\ integrated intensity are shown on the
  SHARC-II 350 \micron\ image (intensities in Jy/beam). Bottom:
  Overlays of \CII\ on 1.47 GHz radio continuum (left, VLA archival data) and
  SHARC-II 350 \micron\ contours (right) on the SUBARU near-IR image.}
\label{cii-co-intint}
\end{figure*}

\section{Results and discussion}
\subsection{Global morphology and the dark lane}
Figure~\ref{cii-co-intint} shows the integrated intensity distribution
of \CII\ 158~\micron\ and CO \eleven\ emission as observed with
SOFIA. The emission morphology in both lines shows a pronounced peak
centered on S106 IR and extended emission towards the south,
north-west, and north-east, creating the heart-shaped emission already
seen in the KAO data. The CO emission is more confined to the central
region, offset from the emission peaks seen in \CII, and better
matches the \OI\ emission from the dense PDR gas \citep{Schneider2}.
The SHARC-II 350 \micron\ data show S106 FIR as a bright, unresolved
source west of S106 IR, while east of the star, the emission is
elongated to the north-east with some fainter extension towards the
south-east resembling a dusty cone. Close to S106 IR, the integrated
\CII\ intensity has two lobes perpendicular to the submm continuum
emission. 


Near-IR and cm continuum show the northern and southern lobe divided
by the dark cone in the east. Near S106 IR, the two \CII\ peaks
closely follow the cm continuum, suggesting that part of the
\CII\ emission is coming from the \HII\ region, while the more
extended \CII\ is unrelated to the cm continuum. The SUBARU image does
not trace the full extent of the \HII\ region just north- and
south-east of S106 IR. Here, there is clearly radio continuum emission
from \emph{behind} the dark cone in the foreground. The submm
continuum emission perfectly matches the dark lane as well as the
fainter near-IR dark regions just south of it. These observations
reveal that the dark lane and cone are \emph{not} just a shadow of the
small disk around S106 IR, but indeed high column density, warm gas at
the edge of the molecular cloud (traced by the --2 \kms\ channel of
\thco\ \two\ in Fig.~\ref{appfig-cii-co-13co21}), located in front of
the \HII\ region. The shadow of the small central disk protects
  the dark lane from ionizing UV radiation and likely is responsible
  for the fact that the lane survived for a significant time of the
  nebula's evolution after the formation of the central star.
We will now use our new SOFIA and complementary IRAM 30 m spectral
line data to study the kinematics of the warm gas in the context of
the nebula structure and geometry. For this purpose, we include
various channel map overlays in the Appendix that reveal how complex
the emission in the different velocity components and for the
different tracers is. We, therefore, focus this letter on the analysis
and discussion of the morphology and kinematics of S106 and defer a
more quantitative analysis, including PDR modelling, to a later paper.

\subsection{Velocity structure}
\begin{figure}[th]
\includegraphics[width=8.5cm]{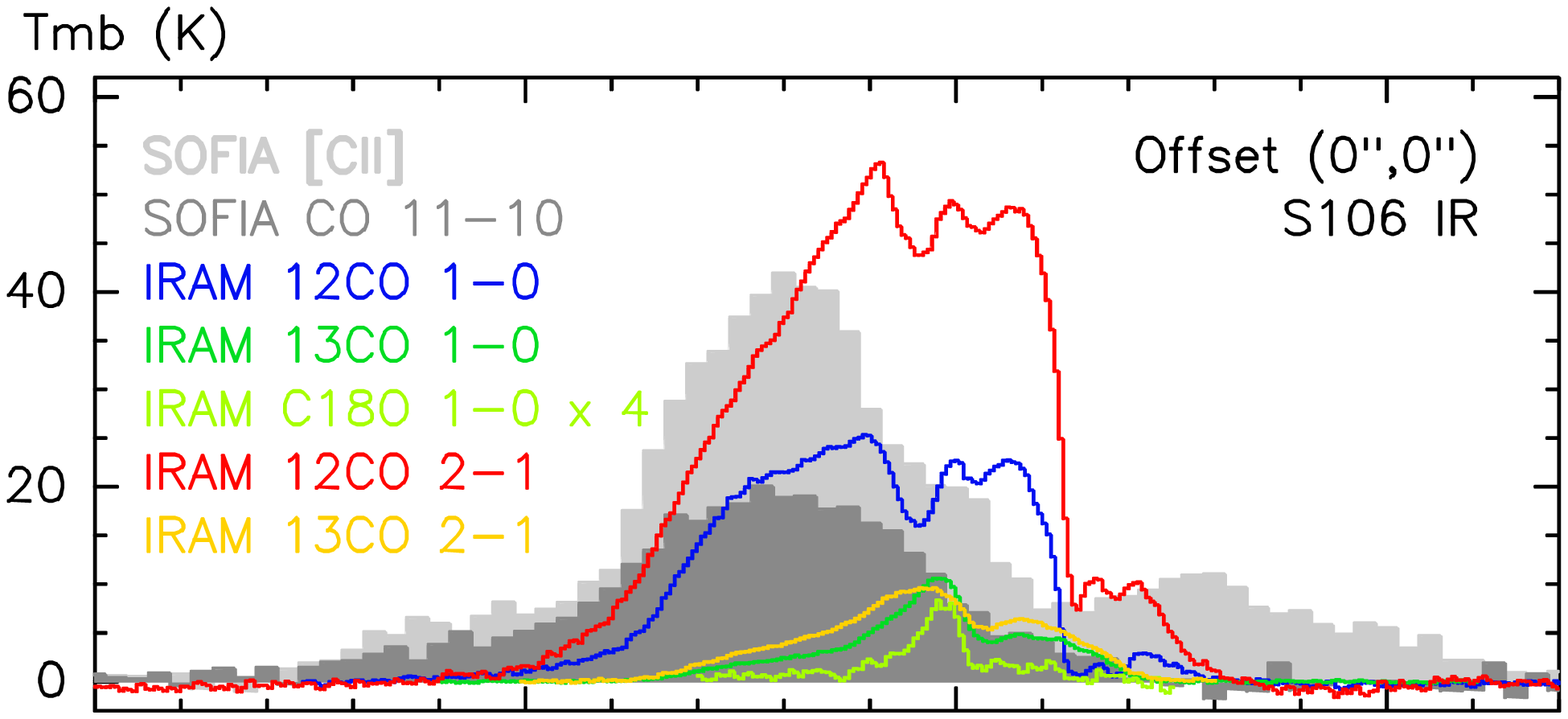}\\
\includegraphics[width=8.5cm]{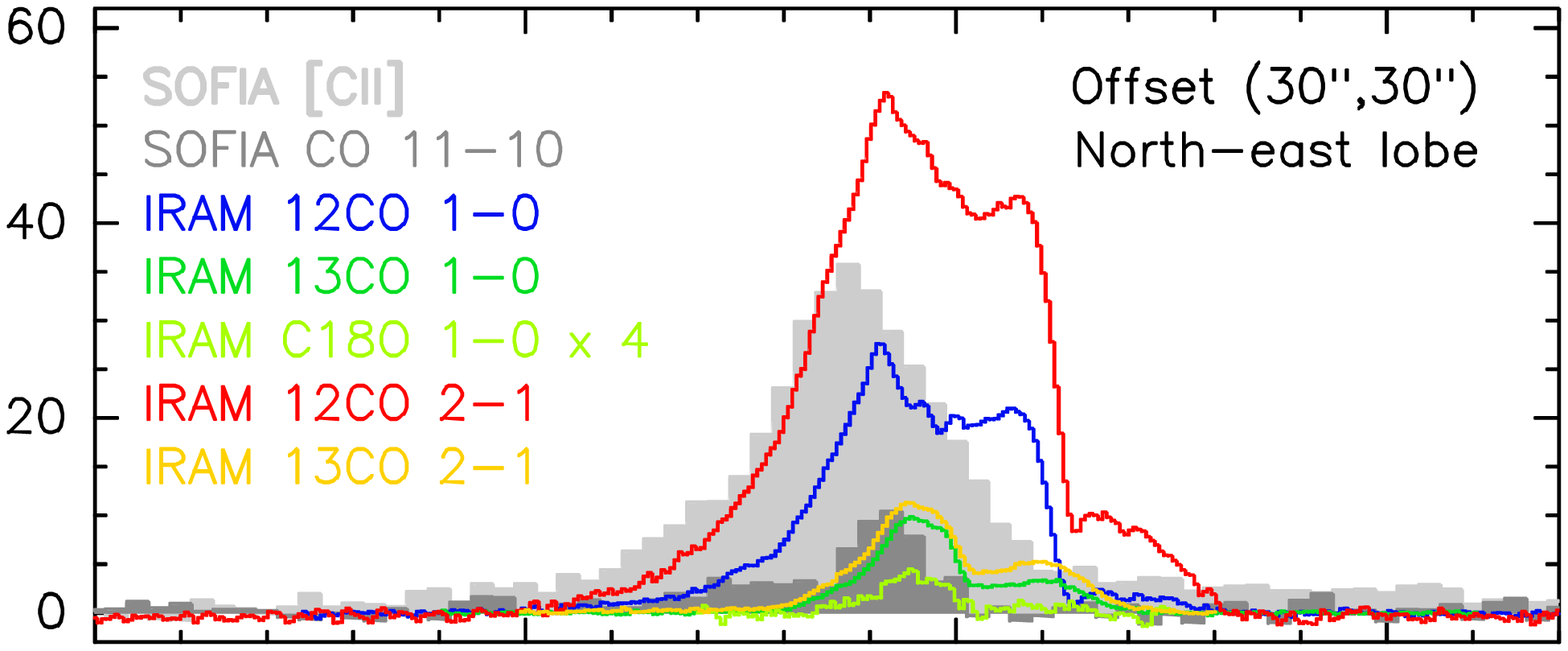}\\
\includegraphics[width=8.5cm]{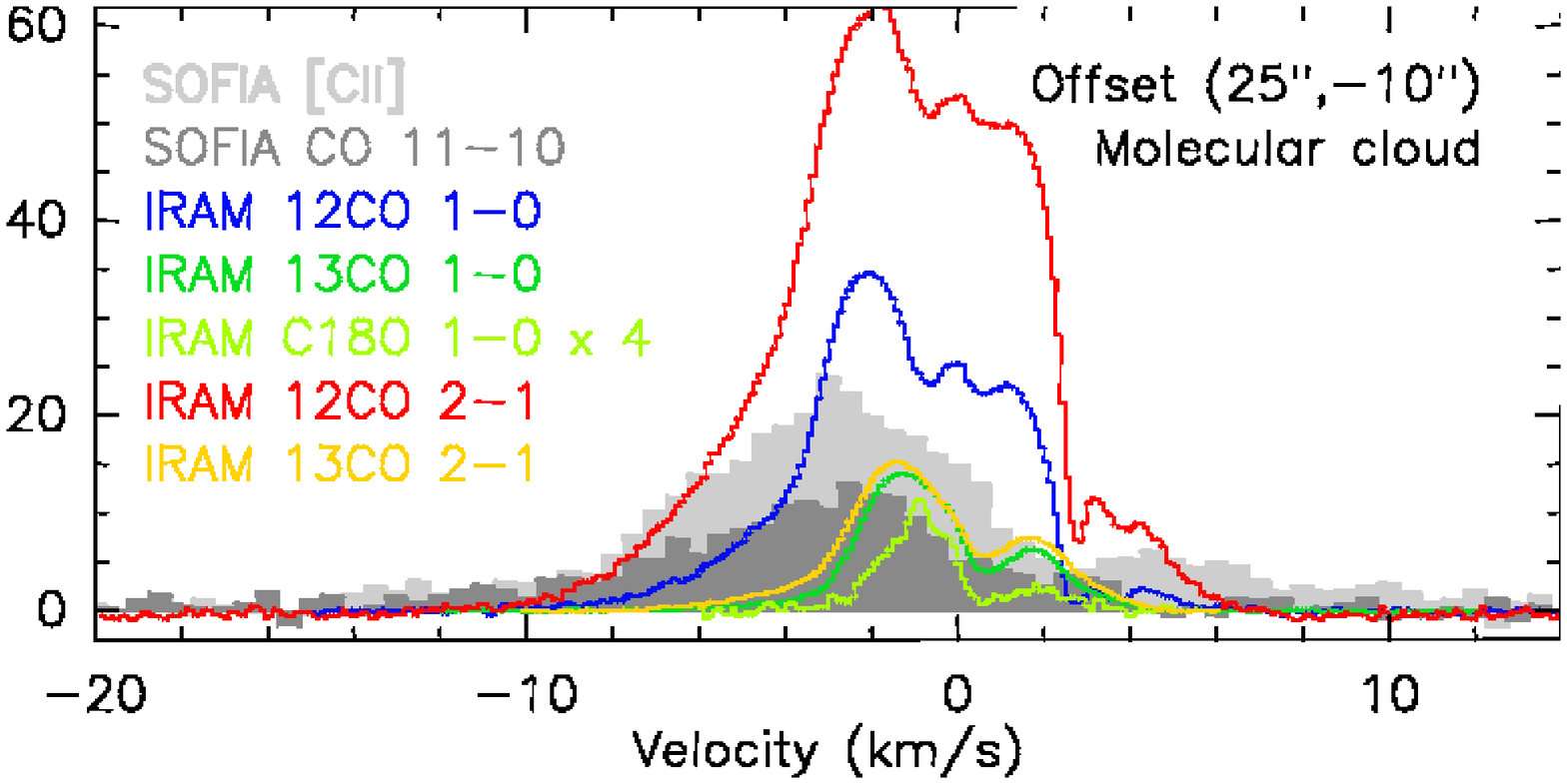}
\caption{\CII\ and CO \eleven\ spectra from SOFIA together with low-J
  CO lines from IRAM \citep{Schneider2} at a common angular resolution
  of 25$''$. Offsets are given in arcsec from S106 IR. From top to
  bottom: S106 IR, north-east lobe, and molecular cloud.}
\label{spectra}
\end{figure}

Channel maps of \CII\ and CO \eleven\ emission (top panel of
Fig.~\ref{appfig-cii-co-13co21}) reveal the different velocity
components of \emph{warm and dense} gas in S106.  The emission can be
divided into three major ranges for \CII: (1) localized \emph{blue}
emission close to S106 IR (v=--6 to --3 \kms), (2) extended emission
with the characteristic heart-shape and an extension to the south at
R.A. offset --70$''$ associated with the velocity of the bulk of
molecular gas (v=--3 to 1 \kms, see low-J CO spectra in
Fig.~\ref{spectra} and channel maps in
Fig.~\ref{appfig-cii-co-13co21}), (3) and faint \emph{red} emission
spatially coinciding with the southern \HII\ region lobe (v=0 to 4
\kms\ and higher). CO \eleven\ is offset from the \CII\ emission in
all channel maps, with \CII\ often wrapping around the CO. These
differences are due to the different excitation conditions. CO
\eleven\ is prominent closer to the star as it traces the warmer,
dense clumps mainly associated with the dark cone close to S106 IR
(the J=11 level is 365 K above ground and the critical density is
$>$$10^6$ \ccm), while \CII\ traces PDR or shocked material at the
interfaces of the cloud and the \HII\ region.

The finding that the northern lobe is dominated by blue and the
southern by red \CII\ emission is surprising as the southern lobe is
tilted towards the observer while the northern lobe is pointing away
(e.g., \citet{Solf1982}). For a proper explanation, we need to
consider what is known about the geometry and dynamics of the nebula.
The lobes of the hourglass nebula expand both radially and along the
main axis \citep{Hodapp2008} driven by S106 IR. The wind and the
expanding lobes are expected to sweep up material where they hit
molecular gas, except for the front wall of the southern lobe which
has been partly eroded by the star (the back side of the southern lobe
is visible in the optical and visual extinction has been shown to be
significantly higher towards S106 IR and the northern lobe
\citep{Felli1984}).  We thus interpret the higher velocity blue and
red \CII\ emission (v$<$--4 and v$>$1 \kms) as arising from swept-up
material at the front and back sides of the expanding, wind driven
hourglass. In this picture, one expects net blue shifted emission
(with respect to the bulk molecular emission at $\sim$--1 \kms) from
those parts of the hourglass close to S106 IR that are curved towards
the observer (front side) and red shifted emission from those curved
away (back side).  Indeed, blue shifted CO and \CII\ emission is
particularly prominent in the northern and southern lobe close to S106
IR where the dark cone converges towards S106 IR in the foreground
(--4 to --2 \kms\ velocity channels).

As most of the front wall along the main axis of the southern lobe has
been eroded, we do not expect extreme blue shifted emission there.
The backside of the southern lobe should also show blue emission
because of its tilt towards the observer. We suggest that this
emission is shifted towards the red due to the radial expansion of the
lobe. Red shifted emission is observed only towards the southern lobe
very close to S106 IR and quite faint further along its axis, which
may indicate that the back side of the northern lobe has also been
eroded.

For emission around the velocity of the bulk of the molecular gas,
\CII\ traces the side walls of the lobes (i.e., where the radial
component of the velocity is smallest).  Comparing \CII\ and CO
\eleven\ emission to $^{13}$CO 2$\to$1
(Fig.~\ref{appfig-cii-co-13co21}) allows to better characterize their
relation to the bulk of the molecular cloud that is well traced in the
low-J CO line.  \CII\ (and to a certain degree CO \eleven) in the --3
to +1 \kms\ range nicely traces the surfaces of the molecular cloud.
We therefore attribute the \CII\ emission in this velocity range as
arising from PDRs in the cavity walls of the ionized lobes, directly
correlated with the bulk emission of the molecular cloud.  Further
support for the above scenario comes from the overlays of
\CII\ velocity channels on continuum data from the VLA (emission from
the \HII\ region) near-IR from SUBARU (emission of hot dust), and
SHARC-II (high columns of warm dust) in Fig.~\ref{appfig-cii-cont}.
The comparison with the 350 \micron\ data is particularly interesting
as it shows that \CII\ is confined to or even funneled into the lobes
by the high column density, warm dust seen in the submm continuum (the
dark cone and S106 FIR). We even observe the dark lane sandwiched
between \CII\ and $^{13}$CO 2$\to$1 emission, showing that there is a
layering of warm ionized gas, warm dust, and colder molecular cloud
material at the surface of the cloud.  The strongest \CII\ emission
originates from the edge of S106 FIR which is probably a signature of
ablation or evaporation of this dense clump due to radiation and/or
wind from the star.

Assuming optically thin, thermalized emission and beam filling of
unity, we estimate lower limits to the beam averaged \CII\ column
density for the emission in the channel maps following
\citep{Crawford1985} in the high density, high temperature limit.  For
the \CII\ wave length of 158 \micron, an integrated intensity of 1 K
\kms\ translates into $7.041 \times 10^{-6}$ \ergbrightness\ and a
column density of $4.6 \times 10^{15}$ \scm. The total \CII\ column
density towards the emission peak is $2.0 \times 10^{18}$ \scm, while
the values for the channel maps in 2 \kms\ bins span a small range
from 1.4 to 2.4 $\times 10^{17}$ \scm, the maximum being in the --4
\kms\ channel. The \CII\ column density in the \emph{red} velocity bins is
of the order of $7.3 \times 10^{16}$ \scm.

\subsection{Spectral signatures of high velocity gas}
The above findings are also reflected in the spectra displayed in
Fig.~\ref{spectra} for three selected positions. The spectra have a
common angular resolution of 25\arcsec\ and show line profiles towards
S106 IR, the north-eastern lobe (cavity wall), and the molecular
cloud. \CII\ emission peaks at velocities of $\sim$--4 \kms, blue
shifted from the peak velocity of the low-J CO lines at --1 \kms, and
is generally very broad with emission covering the range from --14 to
+12 \kms. This implies that \CII\ traces very well the dynamics of
warm gas in the PDR-layers, shocks, or the ionized phase, which most
likely all have slightly different velocities and are subject to
enhanced turbulence, quite different from the bulk of the molecular
cloud.

The CO \eleven\ line peaks near S106 IR at velocities around --4
\kms\ and shows broad lines due to the stellar wind hitting the
molecular gas, thus tracing the warmest part of the molecular cloud
edge in the foreground closest to the star where the wind speed
  is higher or even shocks can play a role. At this position, we
sample emission from both the \CII\ and the CO \eleven\ peak. Towards
the north-eastern lobe, CO \eleven\ becomes much weaker and the peak
velocity shifts towards that of the lower-J lines, likely reflecting
that temperatures and densities are too low for sufficient excitation
in the more quiescent gas outside the \HII\ region. Low-J CO lines
with high to moderate optical depth also show blue wing emission, but
not as extended to the far blue side as observed for \CII\ and CO
\eleven. This confirms that the latter lines trace gas in the
foreground closer to the star and detached from the bulk of the colder
molecular gas.  High velocity red emission above +6 \kms\ is only
prominent in \CII\ and not seen in any of the other tracers.

\section{Summary}
Our new spectrally resolved observations of \CII\ and CO
\eleven\ reveal very complex morphology and kinematics of the warm gas
in S106 with multiple velocity components at different locations
relative to the exciting star, the dark lane, which we show to be
composed of warm, high column density gas, and the bulk of the colder
molecular gas. The \CII\ and high-J CO emitting material is highly
dynamic with broad, non-Gaussian wings in particular in \CII\ showing
no counterpart in any of the other observed tracers for the highest
red and blue shifted velocities.

Only the high spectral and angular resolution provided by GREAT on
SOFIA make it possible to disentangle emission components arising from
the PDR surfaces at the inner cavity walls of the two \HII\ region
lobes, from swept-up gas due to the stellar wind of S106 IR and/or the
expansion of the lobes, and from the \HII\ region itself. To quantify
how much of the emission is contributed by shocks, \HII\ region, or
PDR gas requires a more detailed analysis involving modelling, which
will be addressed in a forthcoming paper.

\begin{acknowledgements}
Based in part on observations made with the NASA/DLR Stratospheric
Observatory for Infrared Astronomy. SOFIA Science Mission Operations
are conducted jointly by the Universities Space Research Association,
Inc., under NASA contract NAS2-97001, and the Deutsches SOFIA Institut
under DLR contract 50 OK 0901. We thank the SOFIA engineering and
operations teams and the DSI telescope engineering team for their
tireless support and good-spirit teamwork, which has been essential
for the GREAT accomplishments during Early Science.
\end{acknowledgements}

\Online

\begin{appendix} 

\section{Channel maps}
\begin{figure}[bh]
\includegraphics[width=16cm]{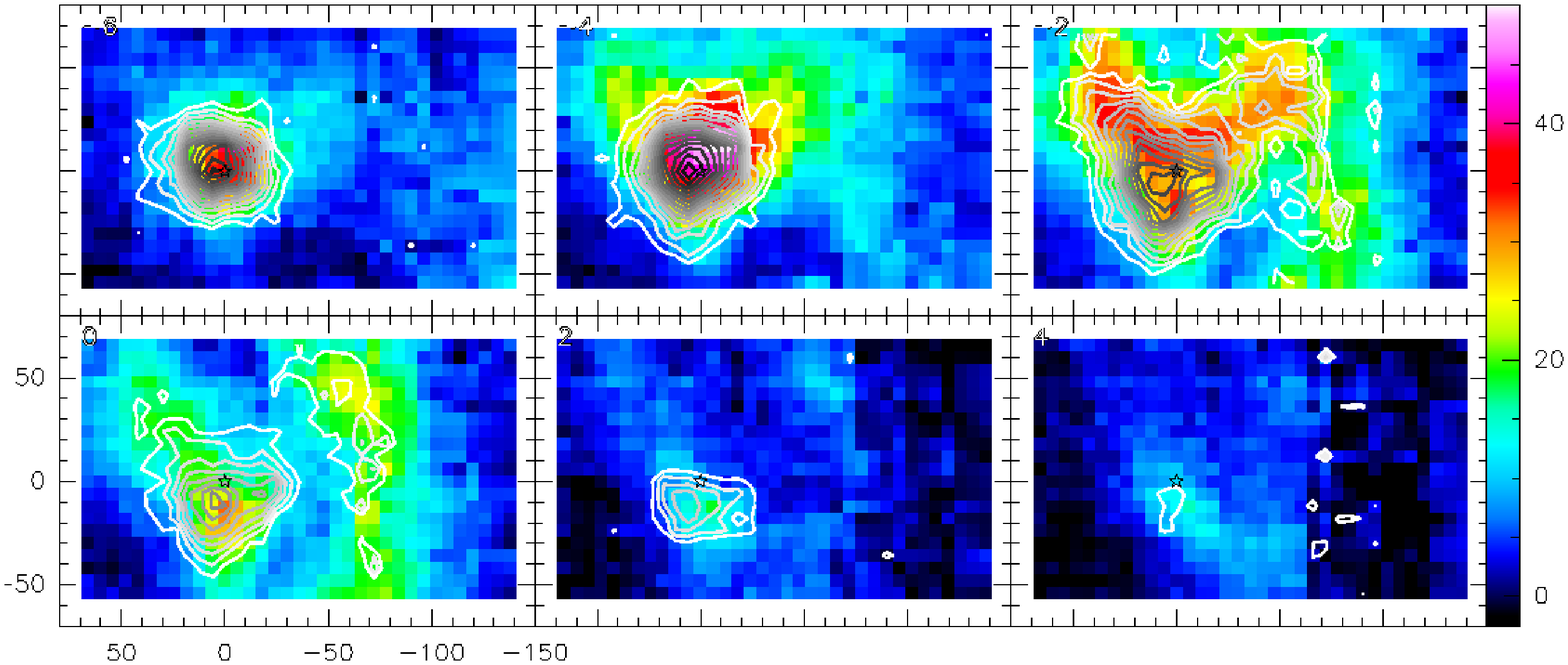}
\includegraphics[width=16cm]{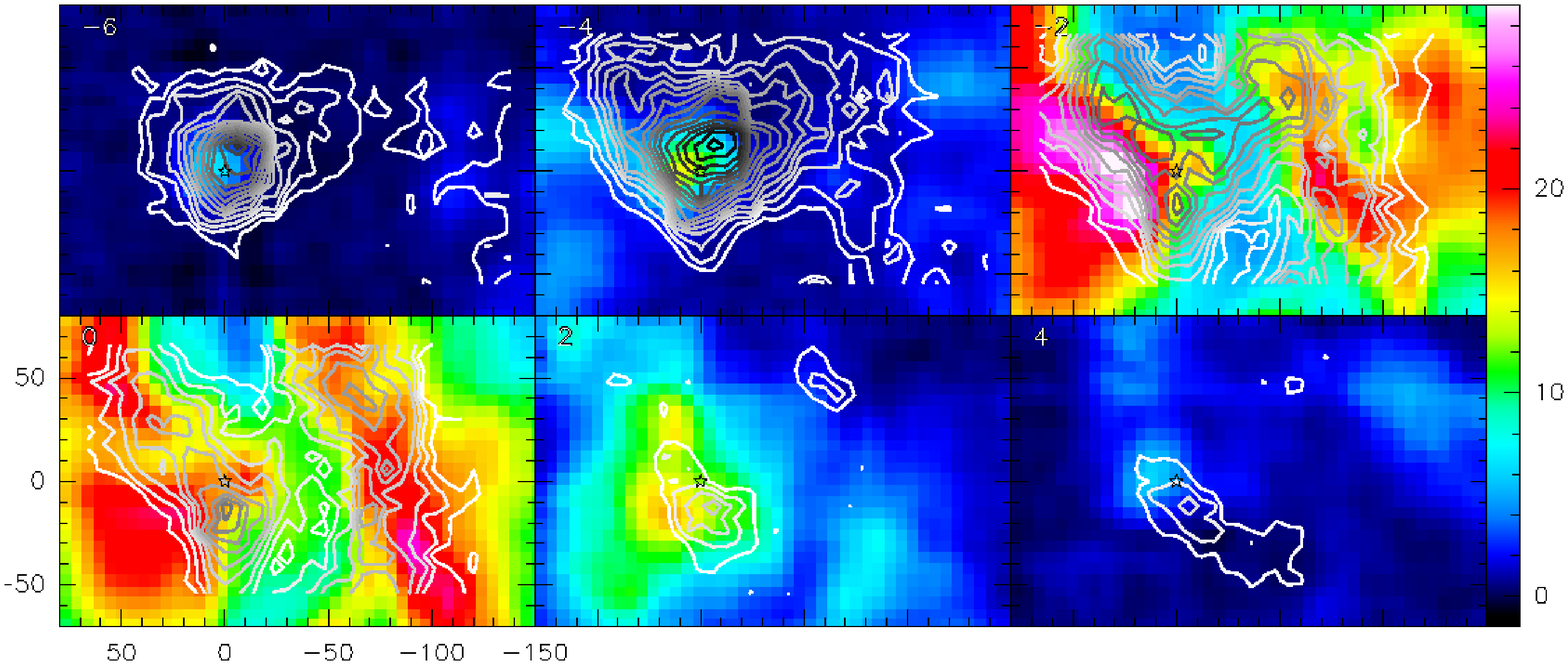}
\includegraphics[width=16cm]{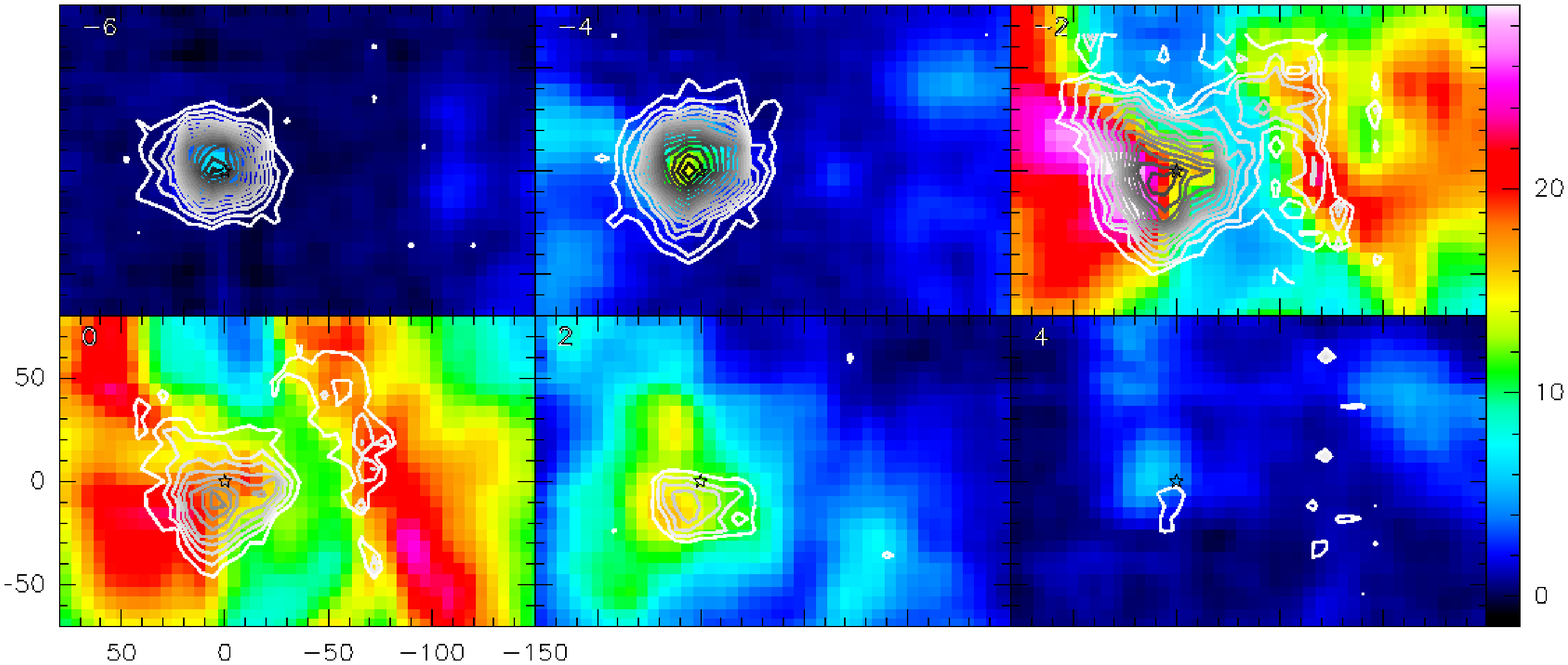}
\caption{Channel maps of CO \eleven\ contours on \CII\ (top) and
  \CII\ and CO \eleven\ contours on \thco\ \two\ (middle and bottom).}
\label{appfig-cii-co-13co21}
\end{figure}

\begin{figure*}[th]
\includegraphics[angle=0,width=16cm]{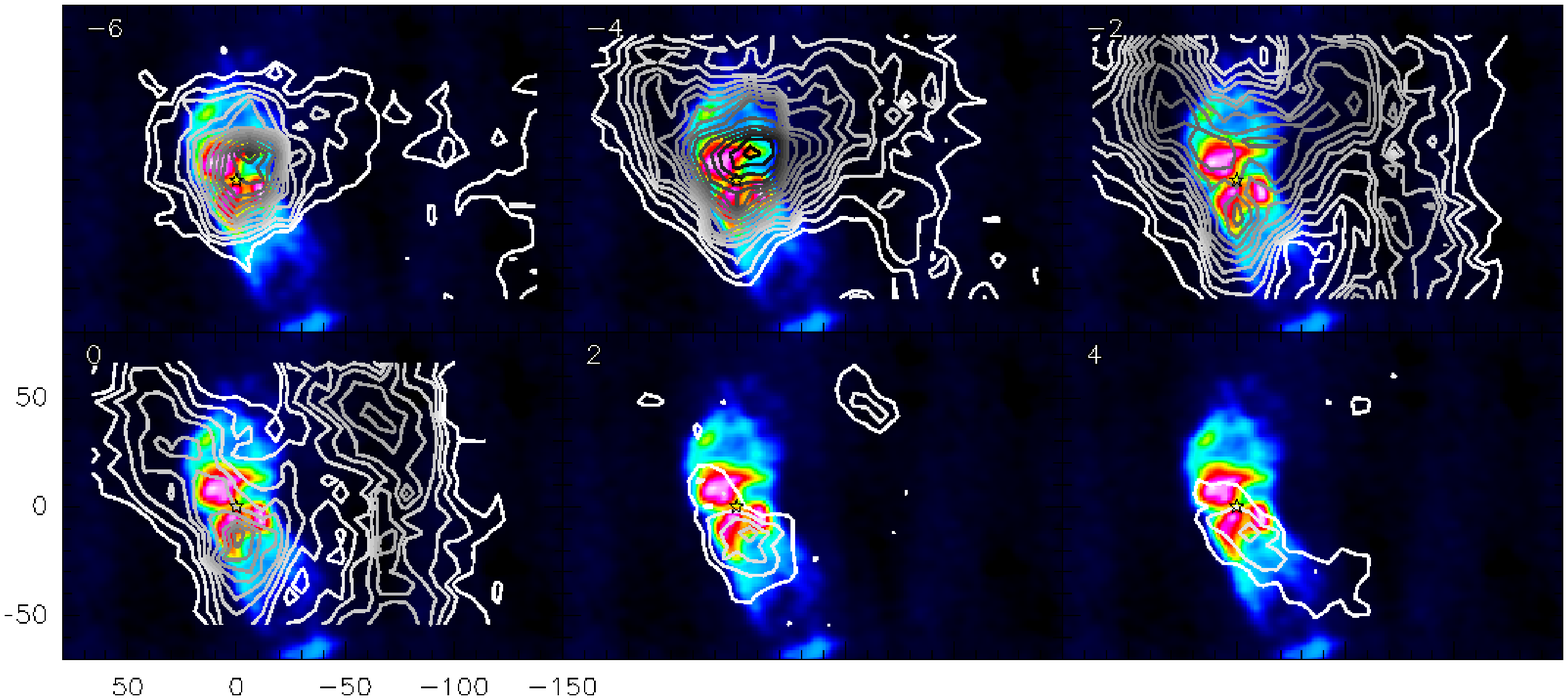}
\includegraphics[angle=0,width=16cm]{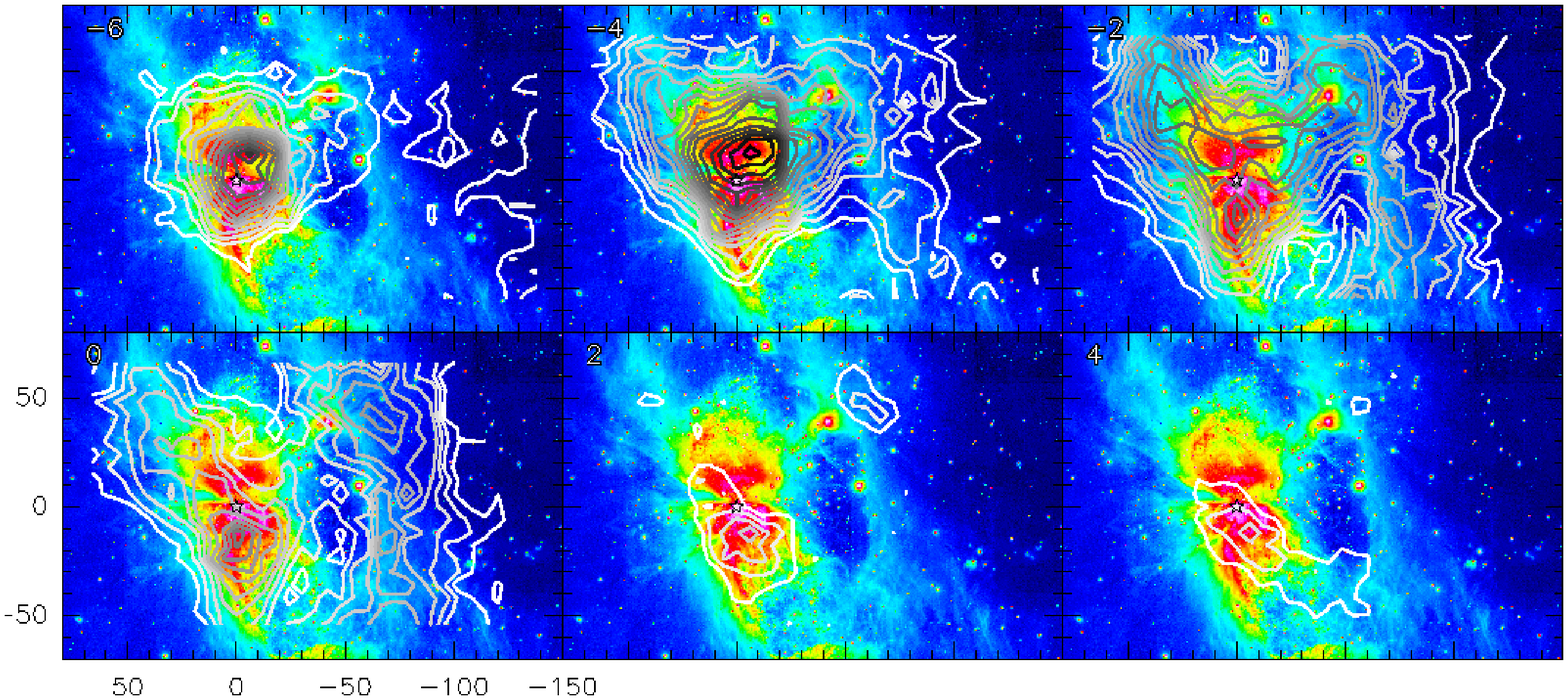}
\includegraphics[angle=0,width=16cm]{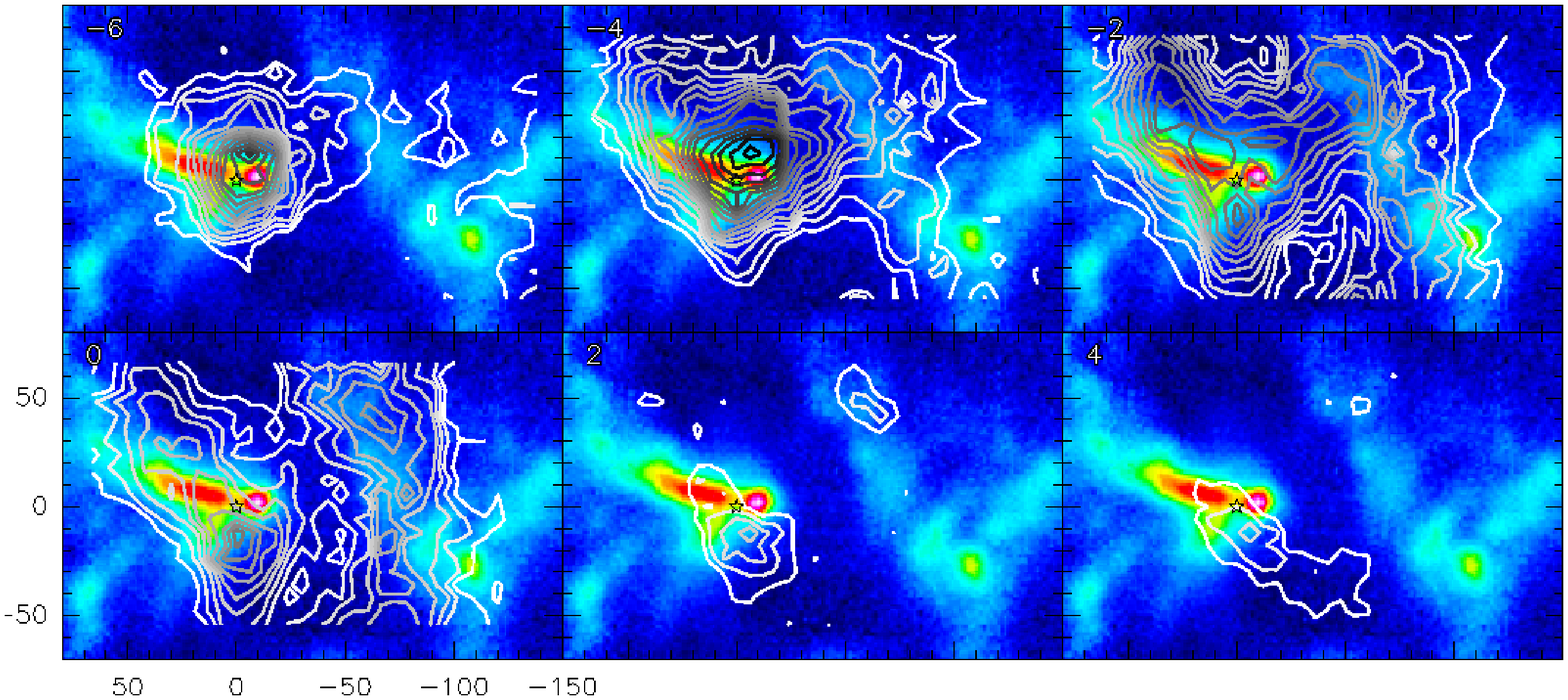}
\caption{Top: channel maps of CII contours on a VLA 1.47 cm image. Middle: channel maps of CII contours on SUBARU near-IR. Bottom: channel maps of CII contours on SHARC-II submm continuum.}
\label{appfig-cii-cont}
\end{figure*}

\end{appendix}


\begin{thebibliography}{}
\bibitem[Bally \& Scoville(1982)]{Bally1982} Bally, J., \& Scoville, N.~Z.\ 1982, \apj, 255, 497
\bibitem[Bally et al.(1983)]{Bally1983} Bally, J., Snell, R.~L., \& Predmore, R.\ 1983, \apj, 272, 154 
\bibitem[Barsony et al.(1989)]{Barsony1989} Barsony, M., Scoville, N.~Z., Bally, J., \& Claussen, M.~J.\ 1989, \apj, 343, 212
\bibitem[Crawford et al.(1985)]{Crawford1985} Crawford, M.~K., Genzel, R., Townes, C.~H., \& Watson, D.~M.\ 1985, \apj, 291, 755 
\bibitem[Felli et al.(1984)]{Felli1984} Felli, M., Massi, M., Staude, H.~J., et al.\ 1984, \aap, 135, 261 
\bibitem[Furuya et al.(1999)]{Furuya1999} Furuya, R.~S., Kitamura, Y., Saito, M., et al.\ 1999, \apj, 525, 821
\bibitem[Gibb \& Hoare(2007)]{Gibb2007} Gibb, A.~G., \& Hoare, M.~G.\ 2007, \mnras, 380, 246 
\bibitem[Graf et al.(1993)]{Graf1993} Graf, U.~U., Eckart, A., Genzel, R., et al.\ 1993, \apj, 405, 249 
\bibitem[Guan et al.(2012)]{Guan2012} Guan, X., et al.\ 2012, \aap, this volume 
\bibitem[Harris et al.(1987)]{Harris1987} Harris, A.~I., Stutzki, J., Genzel, R., et al.\ 1987, \apjl, 322, L49 
\bibitem[Heyminck et al.(2012)]{Heyminck2012} Heyminck, S., Graf, U.~U., G\"usten, R., et al.\ 2012, \aap, this volume
\bibitem[Hoare \& Muxlow(1996)]{Hoare1996} Hoare, M.~G., \& Muxlow, T.~B.\ 1996, Radio Emission from the Stars and the Sun, 93, 47 
\bibitem[Hodapp \& Schneider(2008)]{Hodapp2008} Hodapp, K.~W., \& Schneider, N.\ 2008, Handbook of Star Forming Regions, Volume I, 90
\bibitem[Little et al.(1995)]{Little1995} Little, L.~T., Kelly, M.~L., Habing, R.~J., \& Millar, T.~J.\ 1995, \mnras, 277, 307
\bibitem[Noel et al.(2005)]{Noel2005} Noel, B., Joblin, C., Maillard, J.~P., \& Paumard, T.\ 2005, \aap, 436, 569 
\bibitem[Oasa et al.(2006)]{Oasa2006} Oasa, Y., Tamura, M., Nakajima, Y., et al.\ 2006, \aj, 131, 1608 
\bibitem[Richer et al.(1993)]{Richer1993} Richer, J.~S., Padman, R., Ward-Thompson, D., et al.\ 1993, \mnras, 262, 839 
\bibitem[Schneider et al.(2002)]{Schneider1} Schneider, N., Simon, R., Kramer, C., et al.\ 2002, \aap, 384, 225
\bibitem[Schneider et al.(2003)]{Schneider2} Schneider, N., Simon, R., Kramer, C., et al.\ 2003, \aap, 406, 915
\bibitem[Schneider et al.(2007)]{Schneider3} Schneider, N., Simon, R., Bontemps, S., et al.\ 2007, \aap, 474, 873
\bibitem[Simon \& Fischer(1982)]{Simon1982} Simon, M., \& Fischer, J.\ 1982, \baas, 14, 925
\bibitem[Smith et al.(2001)]{Smith2001} Smith, N., Jones, T.~J., Gehrz, R.~D., et al.\ 2001, \aj, 121, 984 
\bibitem[Solf \& Carsenty(1982)]{Solf1982} Solf, J., \& Carsenty, U.\ 1982, \aap, 116, 54 
\bibitem[Stutzki et al.(1982)]{Stutzki1982} Stutzki, J., Ungerechts, H., \& Winnewisser, G.\ 1982, \aap, 111, 201 
\bibitem[Stutzki \& Winnewisser(1985)]{Stutzki1985} Stutzki, J., \& Winnewisser, G.\ 1985, \aap, 144, 13 
\end{thebibliography}
\end{document}